\documentclass[12pt,a4paper]{article}
\usepackage[english]{babel}
\usepackage{graphicx}

\newcommand{\alt}{\mathbin{\lower 3pt\hbox
   {$\rlap{\raise 5pt\hbox{$\char'074$}}\mathchar"7218$}}}
\newcommand{\agt}{\mathbin{\lower 3pt\hbox
   {$\rlap{\raise 5pt\hbox{$\char'076$}}\mathchar"7218$}}}

\begin{document}
\setcounter{footnote}{0}
\setcounter{equation}{0}
\setcounter{figure}{0}
\setcounter{table}{0}
\vspace*{5mm}

\begin{center}
{\large\bf Computer Model of a "Sense of Humour".\\
I. General Algorithm }

\vspace{4mm}
I. M. Suslov \\
Lebedev Physical Institute of the USSR Academy of Sciences,\\
Leninsky pr., 53, Moscow, USSR\,\footnote{
Present address:

P.L.Kapitza Institute for Physical Problems,

119337 Moscow, Russia

E-mail: suslov@kapitza.ras.ru}

\vspace{4mm}
\end{center}

\begin{center}
\begin{minipage}{135mm}
{\large\bf Abstract } \\
A computer model of "a sense of humour" is formulated. The humorous effect is
treated as a specific malfunction in the processing of information,
conditioned by the necessity of a quick deletion from consciousness
of a false version. The biological function of a sense of humour consists in
quickenning the transmission of processed information into
conscioussness and
in a more effective use of brain resources.
\end{minipage}
\end{center}
 \vspace{5mm}


\begin{center}
{\bf 1. Introduction}
\end{center}
\vspace{3mm}

 In everyday life we use humour for amusement, as "a means of extracting
pleasure from the psychical process" \cite{1} and we never ask
ourselves
why nature has provided us with the sense
of humour. The bare fact that there exists a complex biological
mechanism which causes specific muscular contractions (laughter)
as a reaction to a definite combination of sound or visual images
leads us to conclude that the sense of humour originated at early
stages of the evolution\,\footnote{According to Darwin \cite{2}
antropoid monkeys possess a clearly distinct sense of humour.}
when the possibility of obtaining pleasure should not have of
appreciable importance.  The present paper is an attempt to
answer the question about the biological function of the sense of
humour.

In the proposed scheme, the humorous effect is interpreted as
a specific malfunction in the course of information processing
conditioned by the necessity to delete some information transmitted
to consciousness. The biological function of a sense of humour
consists in  quickening the transmission of processed information into
conscioussness and
in a more effective use of brain resources. The proposed model
accounts for different susceptibility of people to humour,
the absence of a humorous effect from a hackneyed joke, the role
of timing in telling jokes, etc. Some remarks
on other emotions are also given.  In the present work we
formulate a general algorithm for a computer realization of a
sense of humour; in the following paper \cite{11} we discuss the
possible realization of the algorithm in neural networks and the
mechanism of laughter.

\vspace{6mm}
\begin{center}
{\bf 2. Humour from the psychological viewpoint}
\end{center}
\vspace{3mm}

In psychology there exist several viewpoints on humour
 \cite{3,4,5}, the best\,--\,reasoned of which is the
concept of incogruity advanced by the Scotch poet Beattie
\cite{6} in
l776. Its concrete treatments are different in different
investigations; we accept the viewpoint close to the one advanced
in the paper \cite{4}: the humorous effect is a consequence of
the "commutation" of two mutually exclusive images (versions,
estimates) in the human conscioussness. In the simplest cases the
commutation occurs on the level of meanings of a separate word (the play on
words). For example in the joke\,\footnote{\,A detailed
classification of the technical aspects of wit can be found in
Freud's book [1] where examples 2*, 3* are taken from.
We give a simplified classification in  accordance
to the purposes of the present paper; in principle, it embraces
all the cases considered by Freud.}

\vspace{3mm}

(1*) {\it\quad\,\,  "My Uncle William has a new cedar chest"

\qquad\qquad "So! Last time I saw him he just had a wooden leg." }

\vspace{3mm}
\noindent
the word "chest" is at first realized in the meaning of "box" but later it
takes on a meaning of "breast". In other cases the commutation takes place on
the level of more complex images:

\vspace{3mm}

 (2*)  {\it \quad\,\, The horse tradesman: "If you mount this horse
 at 4 in the morning

\qquad\qquad\, then at 7 in the morning you will be at
Pittsburg."

\qquad\qquad The customer: "But what shall I do in Pittsburg at 7
in the morning?"}

\vspace{3mm}
\noindent
Here the words of the tradesman realized as "the characteristic
of horse speed" take on the interpretation "giving directions
how to reach Pittsburg by 7 in the morning". Example

\vspace{3mm}

(3*) {\it\quad\,\, Is this a place where Duke of Wellington said his
famous words?

\qquad\qquad Yes, it is the same place but he never said such
words.}

\vspace{3mm}
\noindent
shows that the commutation may occur along the line of general
estimate of a phrase: the second remark at first gives the
impression of being "natural" or "logical" but later is perceived
as "absurd".

The existence of two incompatible versions we were able to
discover in all jokes;  the explicit "commutation" of versions
takes place in approximately half the cases. The rest of
the jokes are constructed according to the principle
which can be called "the ambiguity scheme". In the example

\vspace{3mm}

 (4*) {\it\quad\,\,  Father (reprovingly): "Do you know what happens
 to liars when they

\qquad\qquad\,  die?"

\qquad\qquad Johnny: "Yes, sir; they lie still."}

\vspace{3mm}

\noindent
the expression "lie still" may be lnterpreted as (1) "be
motionless" or (2) "continue to tell lies". The speciflc feature
of such cases is the practically equal possibility of the two
versions: accordingly there is no definite succession of their
appearance in the consciousness. It may be assumed
that the commutation takes place in such cases also, but
the order in which the versions appear is determined
by random circumstances; repeated commutations are also possible.

The humorous effect is caused not only by "wit" discussed
above, but also by the "comic" (exaggerated movements of a clown,
grimaces, caricatures, parodies and so on). As the main
characteristic of the comic, "the deviation from the norm"
can be accepted;  accordingly, the humorous effect is caused
by the repeated commutations "the norm" ---
"not the norm".\,\footnote{\,Of course,
not every "deviation from the norm"  looks funny;
but it should be taken into account that habitual, oft-repeated
deviations are analogous to hackneyed jokes (see below) and the
weak deviations are easily forced out by other emotions (see the
following paper \cite{11}).}

The laughter from tickling can be connected with the attempt
of the brain to localize the place of irritation of skin; the
result of such localization is invariably rejected because the
irritated place is changed unpredictably (that is the reason
why the tickling should be done by another person).

\vspace{6mm}
\begin{center}
{\bf 3. Information processing  }
\end{center}
\vspace{3mm}

We begin the formulation of the computer model of "a sense of
humour" by analysing information
processing. Suppose that a succession of symbols $A_1$,
$A_2$, $A_3,\ldots$ ("text") is introduced from the outside world
to the brain: it can be a succession of words during visual or
auditory percepption.  In the brain a set of images $\{B_n\}$ is
associated with each symbol $A_n$: for example a set of meanings
(a dictionary family) is put in correspondence to each word. The
problem of information processing  consists in
choosing one image $B_n^{i_n}$ (which is implied in the given
context) from the set $\{B_n\}$. The text will be considered as
"understood" if the succession $B_1^{i_1}$, $B_2^{i_2}$,
$B_3^{i_3},\ldots$ (which visually can be imagined as the
trajectory in the space of images --- see Fig. 1)
\begin{figure}
\centerline{\includegraphics[width=5.1 in]{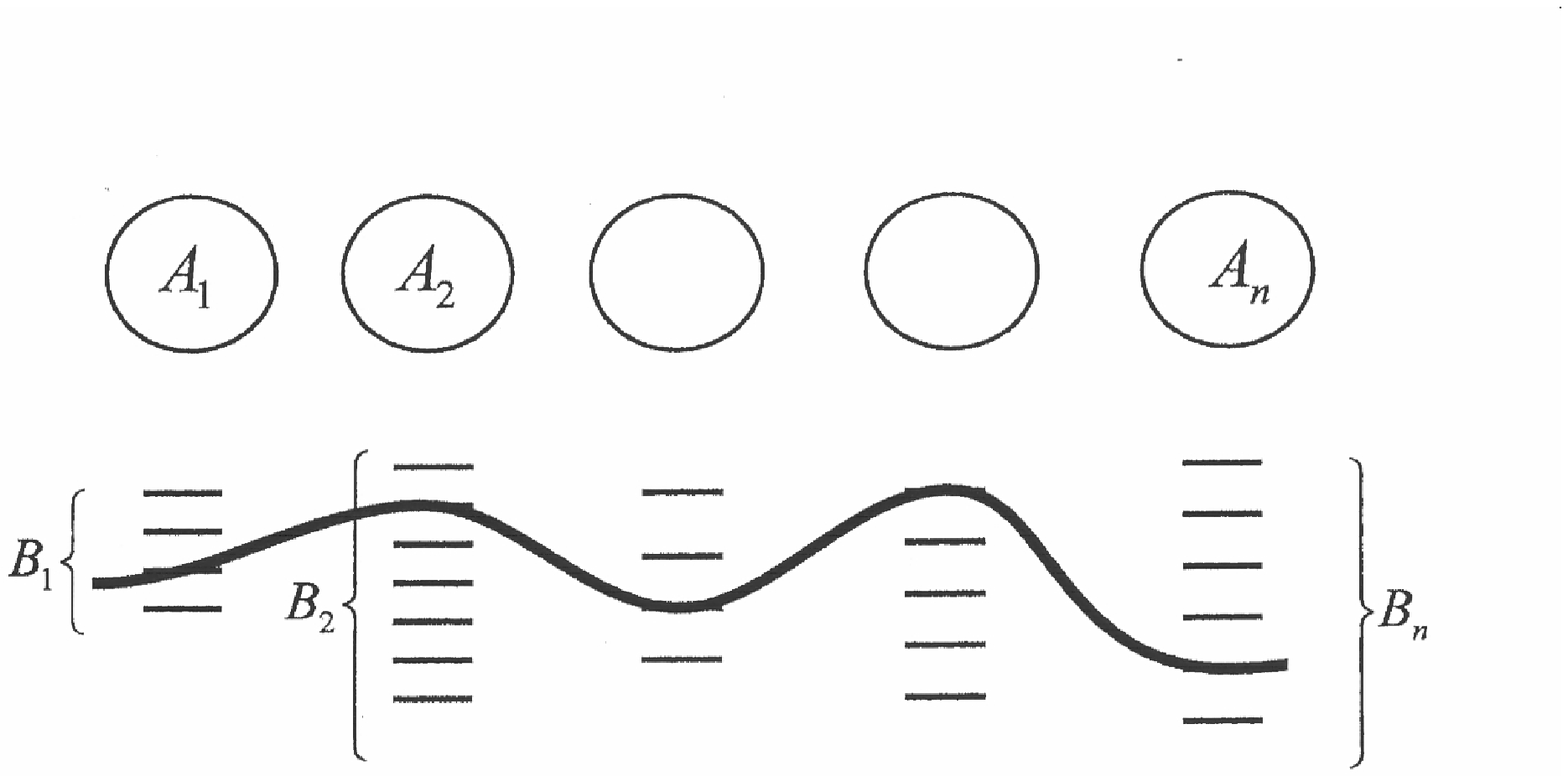}} \caption{The
scheme of information processing: a set of images $\{B_n\}$ is put
in correspondence to  each symbol $A_n$ and  one image $B_n^{i_n}$
should be chosen from the set $\{B_n\}$. The succession
$B_1^{i_1}$, $B_2^{i_2}$, $B_3^{i_3},\ldots$ looks as a
"trajectory" in the space of images. } \label{fig1}
\end{figure}
is put in
correspondence to the succession $A_1$, $A_2$, $A_3,\ldots$ In
principle, the algorithm of information processing
consists in the following:

(1) all possible trajectories in the image space are constructed;

(2) a certain probability is ascribed to each trajectory on the
basis of the information on the correlation of images stored in
memory;

(3) the most probable trajectory is chosen.

Only step 2 is nontrivial here, i.e. the algorithm of
ascribing the probability to a given trajectory. For example,
such algorithm can be based on the binary correlations of images;
in this case the set $p_{ij}$ should be stored in the memory
where $p_{ij}$ is the probablllty of the event that in a meaningful
text image $i$ will be followed by image $j$; the probability
of a trajectory $ijkl\ldots$ is given by the product
$p_{ij} p_{jk} p_{kl} \ldots$. The probabilities
$p_{ij}$ can be obtained by the statistical treatment in the
course of the "learning" process, during which a
sufficiently long fragment of the "deciphered" text (i.e.
recorded in images but not symbols) is introduced to the brain.
A more complex algorithm can take into account the correlation
between $n$ images with $n>2$: then the probabilities
$p_{i_1 \ldots i_{n-1};\, i_n}$ of the succession of images
$i_1 \ldots i_{n-1}$ followed by image  $i_n$ should be
stored in the memory. It is possible to base the algorithm on
binary correlations but with the syntactical connections
taken into account\,\footnote{\,The syntactic structure of
a sentence has a form of a tree, so that each dependent word is
related with its "host". The probability of a trajectory may be
represented as a product of binary probabilities according
to the structure of a syntactical "tree". The practice of machine
translation \cite{7} shows that the syntactic structure in most
cases is clearly established by purely grammatical analysis
(word order, adherence to a part of speach, harmonization of
endings, etc.) and for the purpose of the present work may be
taken as known.}
and so on.
Algorithms of such type are being worked out in the investigations on
machine translation  \cite{7}; the concrete
form of the algorithm is not essential for the following.

The number of operations required for the realization of
any algorithm of such type increases exponentially with the length of the
text. So only fragments of the text containing no more than
a certain number ($N$) of
symbols can be immediately treated by such a method. How can
longer texts be processed? The natural possibility is the
following: during the processing of the first $N$ symbols not
one but several ($M$) of the most probable trajectories are
remembered; then translation on one step made --- the fragment
from the second to the $(N+1)$-th symbols is considered --- and
for each of the $M$ conserved trajectories all possible
continuations are constructed; then again $M$ of the most probable
trajectories are conserved and so on. It is reasonable
to make  the number
$M$ variable, so at each stage as many trajectories
are remembered as the operative memory can hold. In the whole,
the process looks as follows (Fig. 2):
\begin{figure}
\centerline{\includegraphics[width=5.1 in]{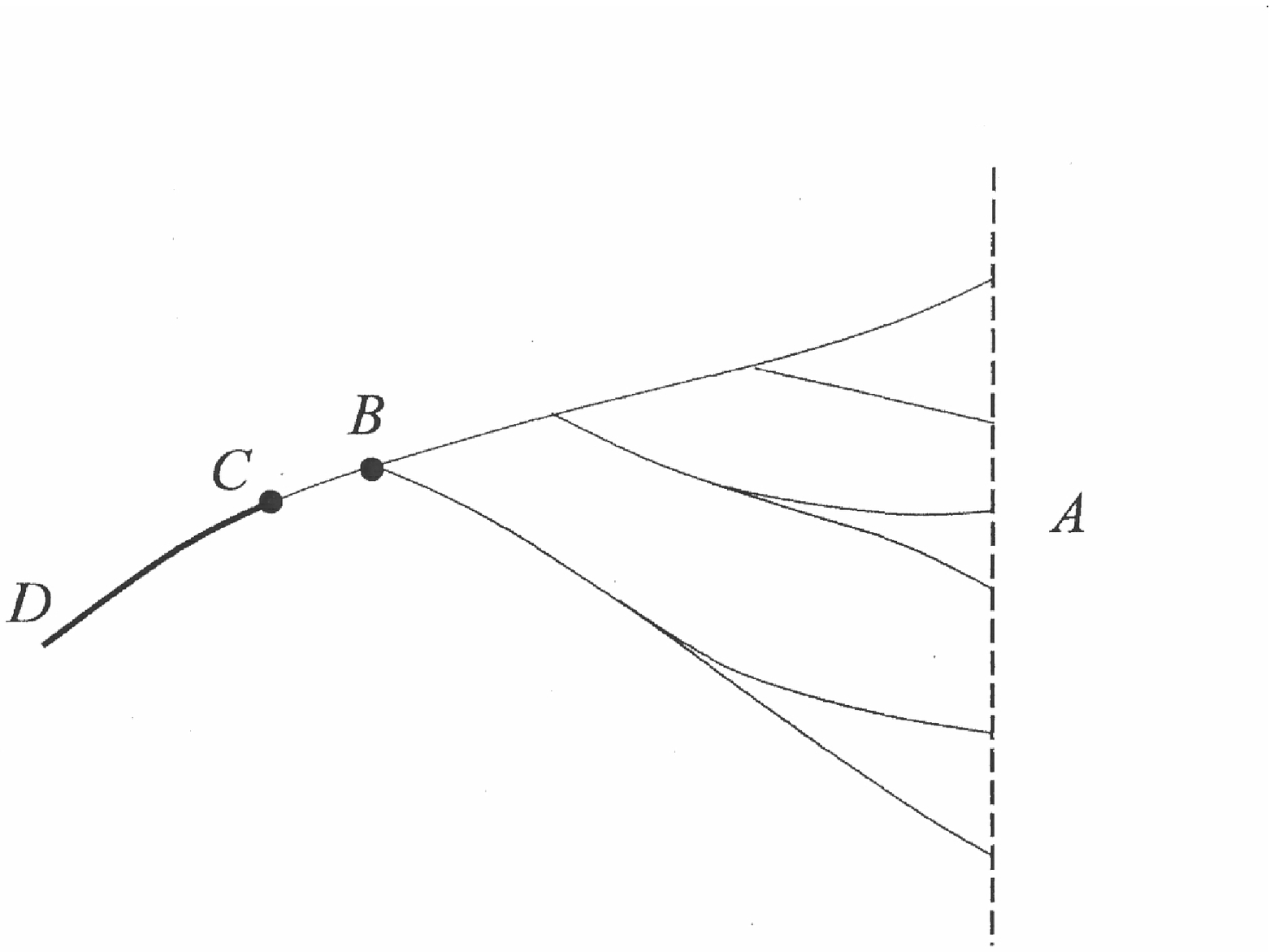}} \caption{The
visual imagination of information processing: thin lines are
trajectories conserved in operative memory, $A$ is a front, $B$ is
the point where the branching is over, $CD$ is a fragment of
deciphered trajectory transmitted to consciousness. } \label{fig2}
\end{figure}
immediately after
the front $A$ the trajectory is branched heavily; at a certain
point $B$ the branching is over (the distance between $A$ and
$B$ is restricted by the volume of operative memory provided
for remembering the trajectories); the deciphered part of
the trajectory $DC$ with some delay $AC$ is transmitted to
the consciousness of the man and is realized by him as a thought
(while the whole process takes place in the subconscious and is
not perceived immediately).

\vspace{6mm}
\begin{center}
{\bf 4. The role of emotions in information processing  }
\end{center}
\vspace{3mm}

If numbers $N$ and $M$ are sufficiently large and the algorithm
of calculating the probability of $N$-symbol trajectory is
good enough, then the described scheme will operate successfully.
However the probabilitistic nature of the algorithm makes
mistakes inevitable: so a mechanism is desirable for minimizing
their consequences. Such mechanism exists and it consists in
communicating to the consciousness some information
about the course of the processing in the subconsciousness; the
man perceives such information as emotions.

For example, such parameters of the process are essential
as the probability $p_{max}$ of the trajectory transmitted
to consciousness and the probability $p_{comp}$ of the most
probable of the competing trajectories.  The high values
of $p_{max}$ and  $p_{max}/p_{comp}$ signal a successful
course of the process and are perceived as positive emotions
(pleasure, confidence): the information obtained is considered
as reliable. The low values of $p_{max}$ and  $p_{max}/p_{comp}$
signal an unsatisfactory course of the
process and  are realized as negative emotions (annoyance, doubt):
the corresponding information should not be taken too seriously.
For very low values of $p_{max}$  no versions are transmitted to
consciousness (complete incomprehension) and so on.

The possible relationship of emotions with the parameters
of the process can be illustrated on the basis of the
semi-empiric "emotion formula" proposed by Simonov \cite{10}
$$
{\cal E}={\cal N}(I-I_0)
$$
where ${\cal E}$ is the emotion strength (which is objectively
measured by the pulse rate, the blood pressure etc.),
${\cal N}$ is a
strength of some need, $I_0$ is the quantity of information
demanded for the satisfaction of this need, $I$ is the quantity
of information the subject has at his disposal (both informations
are estimated subjectively). An emotion is positive
(${\cal E}>0$) for
$I>I_0$ and negative for $I < I_0$.  We can suppose that in the
course of
information processing  ${\cal N}$ is the need in information
and the different parameters of the process determine $I$ and
$I_0$ for different emotions. For example, $p_{max}$  can be used
as $I$ for the emotion "pleasure of understanding --- annoyance of
incomprehension" (accordingly,  $I_0$ is the typical value of
$p_{max}$ ensuring the satisfactory course of the process).
Analogously, $p_{max}/p_{comp}$ can be used as $I$  if ${\cal E}$
is the emotion "confidence\,--\,doubt" and so on.

These speculations lead us to conclude that the emotion
expressing the humorous effect is also related to some specific
situation in the processing of information.

\vspace{6mm}
\begin{center}
{\bf 5. The humorous effect   }
\end{center}
\vspace{3mm}

Let us discuss the nature of the delay of point $C$ with
respect to front $A$ (Fig. 2). At first sight, point $C$ in a
reasonably organized system should be always behind point
$B$ or   coincide with it: it is just the variant we
surely choose writing the computer program. However, for a human
as well as for any living creature such a variant is completely
unsatisfactory. The matter is that the delay of point $C$
 with respect to front $A$ results in the time interval
$\tau_{AC}$  during which the information introduced to the
brain does not appear in the consciousness (the man sees a
bear but he is not aware of this). The prolongation of the
interval $AC$ is  obviously dangerous while the
interval $AB$ can drag out for objective reasons (the man cannot
decide what he sees: a bear or a bush shaped like a bear).
Therefore,
the interval  $AC$ should have the upper bound $\tau_{max}$
on the time scale: if time delay $\tau_{AB}$ corresponding to
the interval $AB$ is less than $\tau_{max}$ then point
$C$ coincides with point  $B$ (Fig.3,a);  if
\begin{figure}
\centerline{\includegraphics[width=5.1 in]{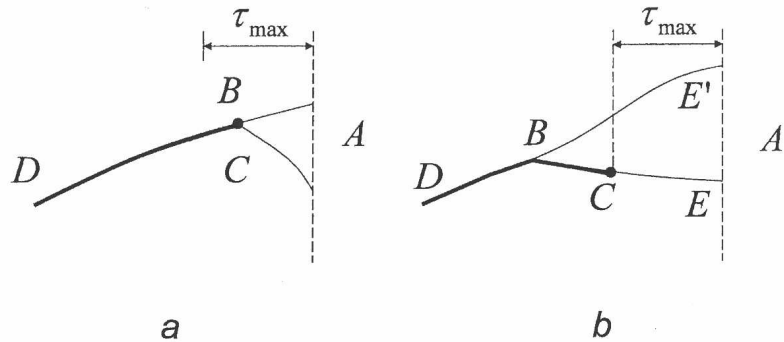}}
\caption{The parameter $\tau_{max}$  is the upper bound of the
time interval corresponding to delay of point $C$ with respect
to front $A$; (a) $\tau_{AB}<\tau_{max}$, (b)
$\tau_{AB}>\tau_{max}$. } \label{fig3}
\end{figure}
$\tau_{AB}>\tau_{max}$, then $\tau_{AC}=\tau_{max}$
and point  $C$ leaves  behind point $B$ (Fig.3,b). In the latter
case, the most probable version $DE$ is transmitted to the
consciousness while competing versions ($DE'$) are conserved in
the operative  memory (Fig.3,b) --- their deletion is unreasonable
because the brain has resources to continue the analysis.
If in the course of the subsequent movement of front $A$
the trajectory $DE$ continues to
have the maximum probability, then the competing trajectory $DE'$
will be deleted and the time will be saved as a result. If in the
course of the movement of front $A$ the probability of $DE$
falls below the probability of $DE'$, then the brain will have
a possibility to correct the mistake. In this case, however,
the specific malfunction occurs: the fragment
$BC$ transmitted to consciousness should be immediately deleted
and replaced by the fragment of trajectory $BE'$.
Psychologically this malfunction is perceived as interference
of two incompatible versions: version $BC$ fixed by the
long-term memory and the newly appeared version $BE'$. The
described specific malfunction can be identified with
"a humorous effect".

Indeed, the situation described is exactly reproduced in
the course of the interpretation of jocular expressions. For
example, in joke (1*) two incompatible versions arise in the
subconsciousness during the analysis of the first remark: in
the first of them ($DE$) the word "chest" is treated as "box"
while in the second ($DE'$) it is treated as "breast". In the
context of the given sentence version $DE$ ("box") is more
probable and is transmitted to consciousness. The appearance of the
word "leg" in the second remark makes version $DE$ less probable
and increases the probability of version $DE'$ ("breast"): this
gives rise to a humorous effect.

It is essential to emphasize that the existence of a humorous
effect is not to any degree unavoidable: nature had a possibility
to avoid it in one of the two manners:
(1) by delaying the transmission of trajectory $DE$ to consciousness
till trajectory $DE'$ is naturally discarded, or (2) by quickening
the transmission of $DE$ by rejection $DE'$ simultaneously.
However, in the first case the time the information reaches
consciousness is delayed and in the second case the brain
resources are not completely used: so nature resolves this
problem at the cost of psychological confusion.

In the process of evolution the optimal value of  $\tau_{max}$
is achieved which ensures the compromise between the reliability
of information and the speed of its obtaining (people with long
$\tau_{max}$ will be eaten by a bear, while people with short
$\tau_{max}$  will confuse every bush with a bear and will be
incapable of getting food). For the optimal value of $\tau_{max}$
the inequality $\tau_{AB}<\tau_{max}$
is satisfied as a rule, and a humorous effect is rare enough
in the natural conditions; but it can be easily produced by
specially constructed witticisms and comics.

\vspace{6mm}
\begin{center}
{\bf 6. Some consequences  }
\end{center}
\vspace{3mm}

The model described offers a natural explanation for a number of
well-known facts.

\vspace{2mm}

{\it The failure of a hackneyed joke to produce a humorous effect}
is a consequence of the fact that a man knows of the existence
of two incompatible versions beforehand and avoids the transmission
of the clearly false version to consciousness (for example,
knowing that in joke (1*) the "chest" turns out to be a "breast"
he is not tempted to interpret it as "box").

\vspace{2mm}

{\it The role of intonation in telling jokes} is related mainly
with temporal characteristics
(pace, arrangement and duration of pauses, etc), which
can be taken into account by incorporating an appropriate number
of "spaces" in succession $A_n$. The quick pace of telling
does not give time for the false version to be transmitted
to consciousness and interval $BC$ (Fig. 3) turns out small or absent.
The slow pace of telling increases the lengths of trajectories
due to "spaces" and the competing trajectory $BE'$ (Fig. 3) is
deleted from the operative memory; so the commutation of versions
becomes impossible.\,\footnote{\,The dependence of humorous
effect on duration of the pause in a certain place is well
described in Mark Twain's essay "Public Speaking". }

\vspace{2mm}

{\it Different susceptibility of people to
humour}\,\footnote{\,We have in mind the susceptibility to humour in
principle, leaving aside the cases when individual peculiarities
give rise to inadequate reaction to a concrete joke. The examples
are incomprehension of a joke due to the absence in memory
of a necessary image, peculiar view of the "norm" while perceiving
the comic, the forcing out of laughter by secondary emotions (see
\cite{11}) and so on.}
 is connected
(in case of equal intellectual level) with the differences in
the delay $\tau_{max}$. People with large  $\tau_{max}$
seldom laugh because point $C$ seldom outruns point $B$. Conversely,
people with small $\tau_{max}$ are aware of a humorous effect
even in cases that most people do not see as funny. Supposedly,
 $\tau_{max}$  is diminished by alcohol and this is
a cause of the unmotivated  gaiety.
At fixed $\tau_{max}$ the susceptibility to humour correlates with
the volume of the operative memory, which determines the average
length of the interval $AB$ (Fig. 2).

\vspace{2mm}

{\it Nervous laughter}. If a mass of unpleasant impressions
rushes at a man and there is danger of the overstrain of the
nervous system then the organism forcibly deletes the unpleasant
information and replaces it by neutral: this gives rise
to the reflectory laughter.

\vspace{6mm}
\begin{center}
{\bf 7. Conclusion  }
\end{center}
\vspace{3mm}

Freud \cite{1} considers the pleasure obtained from laughter as
the main cause of the existence of a sense of humour: a man
discovers the possibility of extracting pleasure from the psychical
process and begins subconsciously and then consciously to exploit
it. Our viewpoint is the opposite: a sense of humour is
biologically conditioned by the necessity to quicken the transmission
of information to consciousness and of a more effective use of
brain resources: so the pleasure obtained from laughter is not
an essential factor (similarly, the two reflexes --- sneezing
and coughing --- exist regardless of the pleasure afforded by the
first and the displeasure caused by the second, because they
are dictated by the biological necessity of cleaning out the
respiratory system). Of course, if laughter afforded displeasure
the social function of humour would change: the society
would try to get rid of it by censorship, prosecution of witty
people and so on.

Is it possible to create a computer program which will "laugh" in the same
cases as a man? From our viewpoint it is quite possible if we restrict
ourselves to the simplest types of jokes based on the commutation of meanings
of separate words (example 1*); the corresponding program will
not be much more complex than the average machine translation
program \cite{7}. Computer modelling of the more complex
jokes involves the need to identify a complete
set of images the average human brain contains and to establish
the correct associative connections between these images.  This
would require many years of work of psychologists and
programmers.

Acknowledgements. I thank L.A. Prozorova for discussion
of linguistic aspects of the paper and D.S. Chemavsky for
discussion of results.

\end{document}